\documentstyle[prb,aps,preprint,epsf]{revtex}
\begin{document}
\draft
\title{Local Density of States in Superconductor -
        Ferromagnetic Hybrid Systems}
\author{Rosario Fazio and Carlo Lucheroni}
\address{Istituto di Fisica, Universit\`a di Catania,
        viale A. Doria 6, I-95129 Catania, Italy\\
        Istituto Nazionale per la Fisica della Materia (INFM), 
        Unit\`a di Catania, Italy}
\maketitle
\begin{abstract}
In this Letter we discuss various properties of the local density of
states (DOS) for a superconductor-ferromagnet hybrid system. The DOS is
modified at small energies on both sides of the interface.  Due to the
interplay of superconductivity and ferromagnetism, the local DOS depends 
on the spin direction. The spin polarization effects extend over a long 
distance from the interface both in the superconductor and in the 
ferromagnet. If the ferromagnet is of finite lenght, the DOS shows a 
(spin dependent) gap. 
\end{abstract}
\pacs{PACS numbers: 74.80.Fp,72.50.Bg, 74.50.+r}
When a normal metal is in contact with a superconductor, pairing correlations 
appear on the normal side. The {\em proximity effect} is intimately related to
the microscopic mechanism which governs the transport through SN interfaces.
An incoming  electron from N is reflected at the interface as a quasi-hole,
and a Cooper pair is injected into the superconductor. Physical
consequences of the proximity effect have been investigated since the early days of  
superconductivity~\cite{deutscher69}.  Nevertheless, thanks to the development of
nanofabrication technology, there has been a tremendous interest in exploring 
mesoscopic effects in hybrid Superconductor-Normal metal (SN) 
hybrid structures~\cite{reviews}.  This activity lead to the discovery of
new phenomena as the magnetoresistance oscillations in SN
system~\cite{petrashov93} and the reentrant effect~\cite{charlat96,stoof96}.    

Due to the proximity effect, there is a modification of the  local density of
states (DOS) as it was recently shown both experimentally~\cite{gueron96} and
theoretically~\cite{golubov88,belzig96}. The local DOS in SN systems depends on the
position relative to the interface. In the normal metal the DOS is suppressed 
at low energies whereas in the superconductor new states appear below 
the BCS gap. Properties of level statistics in hybrid system have been studied in 
Refs.~\onlinecite{frahm96,altland98}. 

The proximity effect is very sensitive to the presence of a strong 
electron-electron interaction, as in the case of nanostructures~\cite{fazio97}, 
and to pair-breaking effects. In this Letter we focus on this last aspect by
considering the case in which the metal is an itinerant ferromagnet.
Properties of ferromagnetic-superconductor (FS) hybrid structures have been
intensively investigated in the past~\cite{bulaevskii85}. It has been shown, 
for instance, that in FS bilayers and multilayers~\cite{radovic91,demler97} 
the critical temperature and critical current oscillate as a function of the 
exchange field and thinkness of the ferromagnet. 
Both effects are related to the center of mass momentum aquired by the 
Cooper pair due to the Zeeman splitting. Further consequences of this phase shift 
can be observed in mesoscopic structures where the Andreev reflected  hole retains 
its phase coherence with the incident electron~\cite{dejong95,upadhyay98}. 
Recent experiments suggest long-range coherence effects also in FS 
structures~\cite{longrange} and show a pronounced reentrant effect  in 
transport for a ferromagnetic wire in contact with a superconductor~\cite{giroud98}.
Furthermore the presence of the exchange field leads to additional structure in the 
subgap current-voltage characteristics~\cite{leadbeater98}.

\begin{figure}
\centerline
{\epsfxsize=11cm\epsfysize=9cm\epsfbox{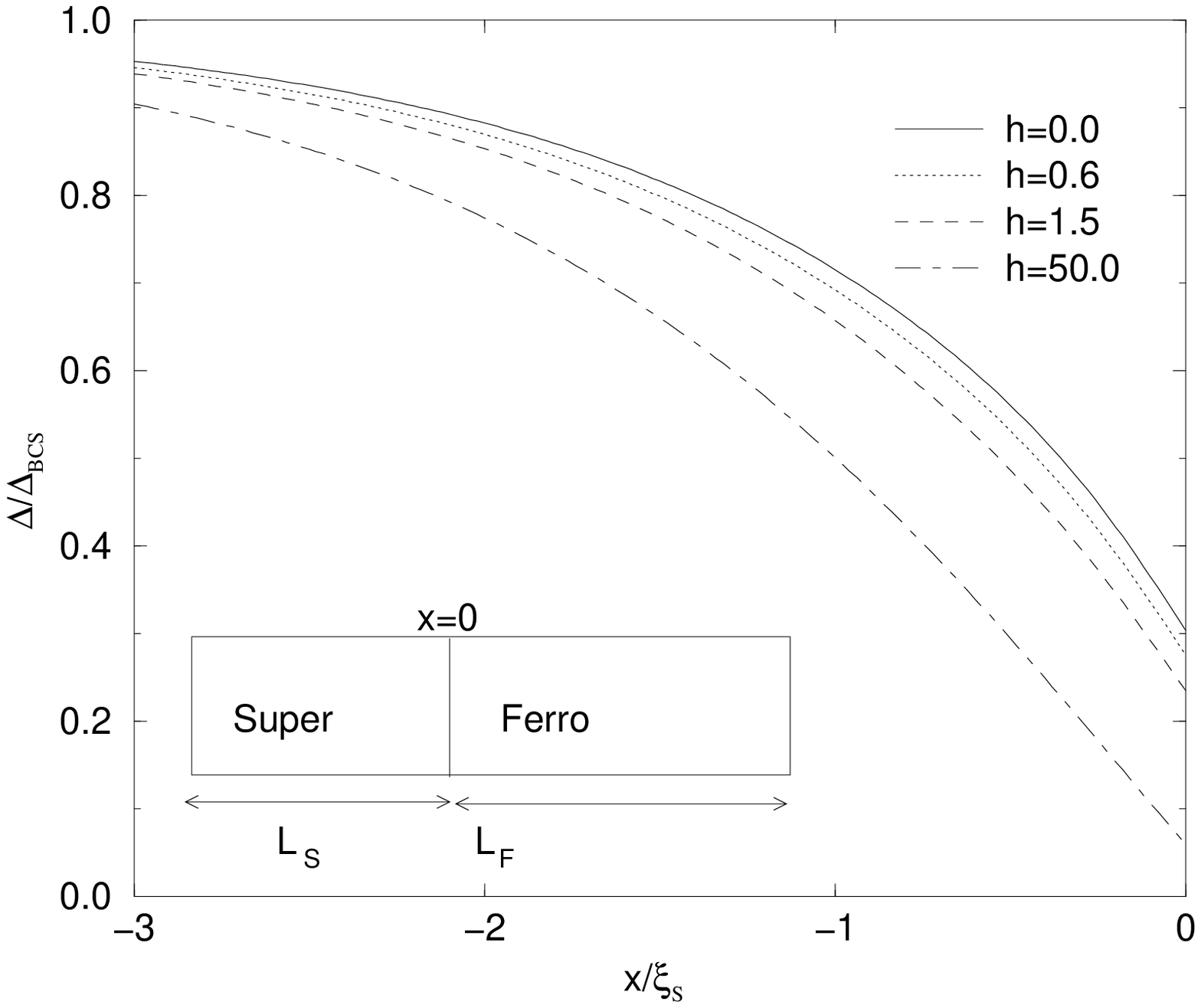}}
\caption{The self-consistent gap is shown as a function of the distance
        from the interface for various values of $h$.
        The system has dimensions $L_F =L_S = 10 \xi _S $.
        Inset: the hybrid FS system considered in
        this work. On the l.h.s. the superconductor, on the r.h.s. the
        ferromagnet.}
\label{fig1}
\end{figure}

An experiment which measures the local DOS in an FS system similarly to that
of Ref.\onlinecite{gueron96} seems to be feasable and it may lead, in our
opinion, to further understanding of proximity in ferromagnetic materials.  
In this Letter we therefore study this problem by solving self-consistently the 
Usadel equations for an FS interface. We will show that the combined effects of
superconductivity (which introduces an energy dependence in the DOS) and
ferromagnetism (which results in a finite momentum for the Cooper pairs
entering the ferromagnet) leads to a rich structure in the DOS. The effects
discussed in this Letter occur when the exchange energy is of the order of the
superconducting gap. There is also a {\em magnetic proximity} effect which, 
for instance, induces a spin dependent DOS also on the superconducting side. 
Zeeman splitting of the DOS  of magnetic materials and superconductors has 
been intesively investigated~\cite{meservey94}. Most of the attention 
was devoted to tunnel junctions. In this Letter we concentrate on the 
opposite limit, a highly transparent FS bilayer shown schematically in 
the inset of Fig.~\ref{fig1}. 

In the absence of an applied magnetic field the physical quantities depend 
only on the relative distance $x$ from the interface. Superconductor and 
ferromagnet have finite length, $L_S$ and $L_F$ respectively. The ferromagnet 
is described by the Stoner model which consists of an effective one-electron 
Hamiltonian in an exchange field. 
For simplicity we assume that the only effect of electron-electron 
interaction is to  create the exchange field and therefore we ignore any 
residual interaction which can induce a finite superconducting gap on the F 
side.  

In the dirty limit, the quasiclassical retarded 
Green's function $\hat{g}(x,E)$ satisfies the Usadel equation~\cite{schmid81}
\begin{equation}
        D \; \partial _x \hat{g} \cdot \partial_x \hat{g}
        + i(E + \mu _B H(x))
        \left[\hat{\tau} _{z}, \hat{g} \right] +
        \left[\hat{\Delta }(x), \hat{g}\right] =0
\label{usadel}
\end{equation}
with the constraints $\hat{g} \hat{g} = 1$ and $\mbox{Tr} \hat{g} = 0$. The hat
refers to the the Nambu notation ($\hat{\tau}_z$ and $\hat{\tau}_y$ 
are the Pauli pseudospin matrices) . 
In eq.(\ref{usadel}) $D$ is the 
diffusion constant, $\mu _B$ is the Bohr magneton, $H(x)$ is a magnetic 
exchange field (we assume that all the material parameters are identical on 
both sides of the interface).
The gap matrix $\hat{\Delta}(x) = \Delta(x) \; \hat{\tau}_y\;$ is determined 
self-consistently ( $\Delta(x)$ can be chosen to be real) by means of 
\begin{equation}
        \Delta(x) = 
        \lambda(x) \int^{\omega_D}_{0}dE\; {\rm Im} [g_{12}(x,E,H)+
        g_{12}(x,E,-H)] 
        \; \;\; \; 
\label{selfconsistent}
\end{equation}
where $\lambda(x)=\lambda \Theta (-x)$ ($\lambda= \mbox{arcosh}(\omega_D /  
\Delta_{BCS})$ is the BCS coupling constant, 
$\Theta(x)$ is a step function), $\Delta_{BCS}$ the BCS gap and $\omega_D$ is
the Debye cutoff frequency. The exchange field $H(x)=H \Theta (x)$ is not
determined self-consistently, hence we do not consider the possible
modification of the domain structure in the ferromagnet due to the presence
of superconductivity~\cite{buzdin88}.
In the problem under consideration, there are two important energy scales, 
the Zeeman energy $\mu _B H$ and the BCS gap $\Delta _{BCS}$. Associated 
with them there are two typical length scales $\xi_F = \sqrt{D/2\mu _B H}$ 
(the coherence length in the ferromagnet) and  $\xi_S = \sqrt{D/2\Delta 
_{BCS}}$ (the superconducting correlation length). 
Since we study the case of perfect transmission at  the FS interface 
the Green's functions and their derivatives are continuous at $x = 0$.

The local DOS $N_{\uparrow (\downarrow)}$ for spins parallel ($\uparrow$) and 
antiparallel ($\downarrow$) to the exchange field $H$ are
\begin{equation}
        N_{\uparrow (\downarrow)}(x,E)  = 
        \pm N_0\, {\rm Re} \, g_{11,(22)}(x,\pm E,H) 
\label{dos}
\end{equation}
where $N_0$ is the DOS (per spin) of a normal metal at the Fermi surface.
We solve the problem at zero temperature. Then, as long as the temperature is
kept well below the critical temperature of the superconductor, all our
results (possibly smeared by  inelastic scattering) are valid.  

The scattering at the FS interface causes pairbreaking~\cite{tokuyasu88}. 
The gap $\Delta$ is suppressed close to the interface more strongly than in 
to the SN case. In the ferromagnet the pair amplitude 
$ 
F(x) = \langle \psi_{\uparrow} (x) \psi_{\downarrow} (x) \rangle
$ 
will be non-vanishing. In Fig.~\ref{fig1} the self-consistent gap is 
calculated for different values of the dimensionless exchange field 
$h=\mu _B H /\Delta_{BCS}$.
By increasing $h$, the value of $\Delta$ close to the interface is 
suppressed and it almost vanishes for the largest exchange field
which we consider. The scale at which the superconducting gap reaches the 
bulk value is $\xi_S$ (independent of the exchange field).
It is possible that for  stronger exchange fields (we are currently 
investigating this issue), $\Delta(x)$ may show a nontrivial behaviour  
close to the interface. Indeed this would not a new situation. A
superconductor in a uniform exchange field may have different forms of 
pairing. In the Larkin-Ovchinnikov-Fulde-Ferrell state~\cite{loff}, for 
instance, the gap is modulated with a period proportional to the exchange 
field.

\begin{figure}
\centerline
{\epsfxsize=11cm\epsfysize=9cm\epsfbox{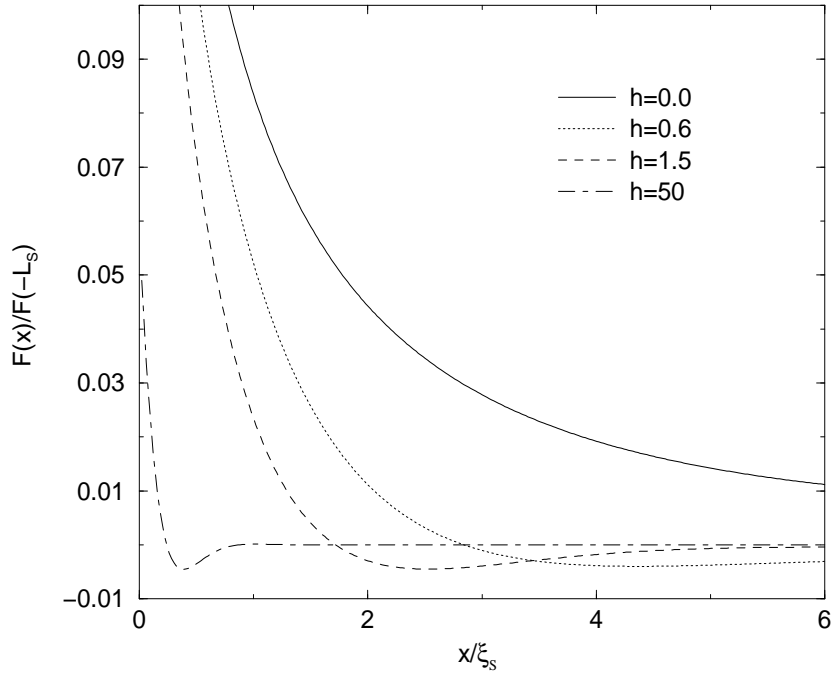}}
\caption{Pair amplitude on the ferromagnetic side for several values of the
        exchange field ($L_S = 10$ and $L_F = 10$).}
\label{fig2}
\end{figure}

For a Cooper pair entering the ferromagnet, the electron and the hole
states are shifted by the Zeeman energy. As a consequence, the pair
amplitude (i.e. the Cooper pair density)  will show rapid oscillations on
the scale of $\xi _F$~\cite{demler97}.  
In Fig.\ref{fig2} we show the results of our self-consistent
calculation of the pair amplitude. On incresing the exchange field, the
pair amplitude drops faster to zero and the period of the oscillation
becomes shorter. Only the first oscillation is clearly seen
while the others are strongly damped due to pairbreaking effects.

We now turn to a detailed discussion of the local DOS. The most important  
findings of Ref.\onlinecite{belzig96} can be summarized as follows. 
i) The DOS in the normal and in the superconducting side of the system aquires a
nontrivial dependence on the energy: the superconductor has a finite DOS
below the gap while the normal metal has strong suppression DOS at the Fermi energy.
ii) If the normal metal has a finite length $L$ there is a minigap in the
excitation spectrum which scales as $\sim L^{-2}$.  The value of the minigap
does not depend of the distance from the interface.
iii) The distance at which the anomaly in the DOS vanishes is energy dependent
and it goes as $\sim \sqrt{D/2E}$. 

\begin{figure}
\centerline
{\epsfxsize=11cm\epsfysize=9cm\epsfbox{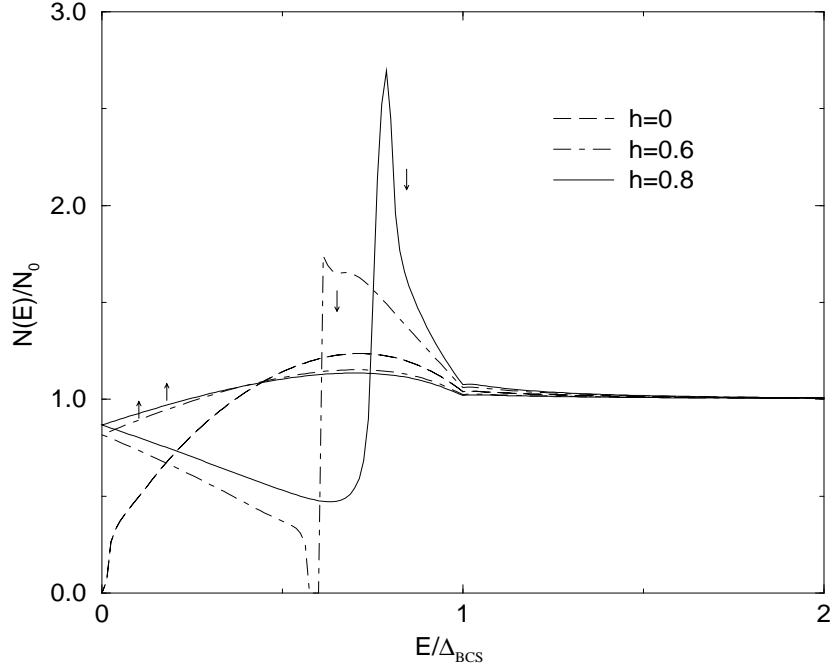}}
\caption{DOS for the system with $L_S =10$, $L_F =10$ on the
        ferromagnetic side, close to the interface ($x=0.1 \xi_S$), for
        different values of the exchange fields, ($\uparrow / \downarrow$) 
        refer to parallel and antiparallel spin respectively. 
        Compared to the normal metal ($h=0$), the anomalies in the 
        DOS are enhanced by the ferromagnetic correlations. }
\label{fig3}
\end{figure}

\begin{figure}
\centerline
{\epsfxsize=11cm\epsfysize=9cm\epsfbox{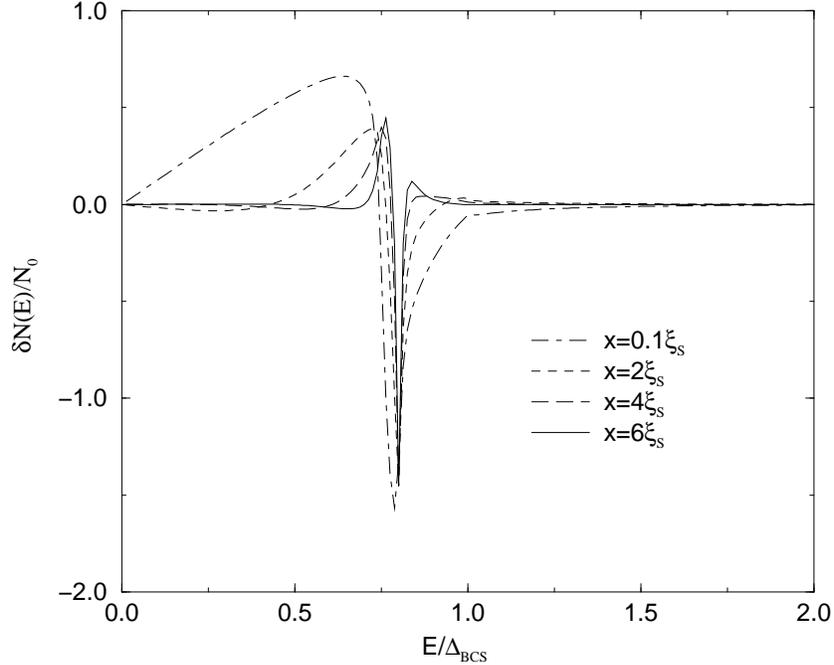}}

\caption{The different in the DOS for opposite spin directions $\delta N$,
        at different distances from the interface on the F side ($L_S =10$,
        $L_F =10$, $h=0.8$).}
\label{fig4}
\end{figure}

All these effects are seen also in FS hybrid systems, with quantitative 
differences though. Here we concentrate on those properties of the DOS which
depend  on the presence of the intinerant ferromagnetism on the normal
side. In particular  we discuss the dependence of the DOS 
on the exchange field and the spin polarization effects. 
The DOS in the ferromagnet is shown in Fig.~\ref{fig3}. Its behaviour
for spin parallel and antiparallel to the exchange field is very
different. Whereas for parallel spin the anomaly is strongly suppressed by the
field, it is  shifted to an energy  of the order of $\mu
_B H$ for antiparallel one.  The anomaly is enhanced compared to the normal
system ($H = 0$) (similarly to what observed in spin polarized tunneling 
experiments~\cite{meservey94}). In Fig.~\ref{fig3} we chose $x=0.1\xi_S$. 
We stress, however, that the anomalous behaviour of the DOS is long-range and
persists over many correlation lengths.
The anomaly is strongest when $\mu _B H \sim \Delta _{BCS}$, by further 
increasing $H$ it progressively shrinks and the DOS approaches a constant.
Even if pairing correlations die out on scales of the order
of  $\xi _F $, the anomalies in DOS survive over a much longer
length scale. The different behaviour of spin up and spin down electrons
on the F side (the spin polarization effect) is shown in Fig.~\ref{fig4} where 
$\delta N =N_{\downarrow} (x,E) - N_{\uparrow} (x,E)$ is analyzed as a 
function of energy and position. The spin polarization of the DOS is 
{\em entirely} due to the proximity effect. Due to proximity the DOS aquires 
an energy dependence also on the normal side of the interface. The Zeeman 
splitting, due to the exchange fiels on the F side, leads to a different 
behaviour for the two spin directions. 

\begin{figure}
\centerline
{\epsfxsize=11cm\epsfysize=9cm\epsfbox{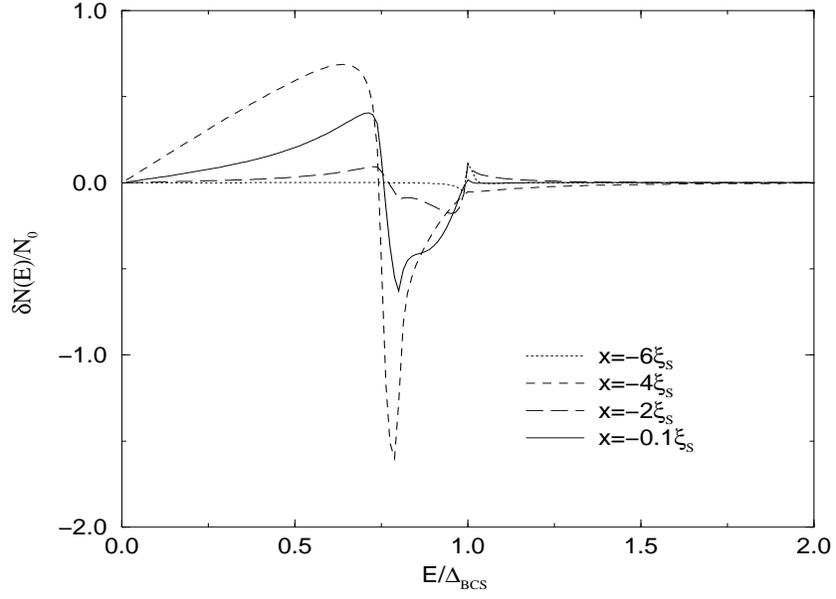}}
\caption{The same as in Fig.\protect{\ref{fig4}} on the S side ($L_S =10$,
        $L_F =10$), $h=0.8$.}
\label{fig5}
\end{figure}

There is also a {\em magnetic proximity} in the superconductor. A finite 
exchange field is induced close to the interface, up to distances of the
order of the superconducting correlation length~\cite{tokuyasu88}. 
Here we emphasize another aspect of this
proximity which occurs in the superconductor (see Fig.~\ref{fig5}). 
Both in the F and S sides, spin polarization effects are most pronounced at
energies of the order of the Zeeman splitting, and extend over many coherence
lengths far from the interface. 

\begin{figure*}

\centerline
{
\begin{minipage}{4cm}
\epsfysize=6cm
\epsfxsize=8cm
\epsfbox{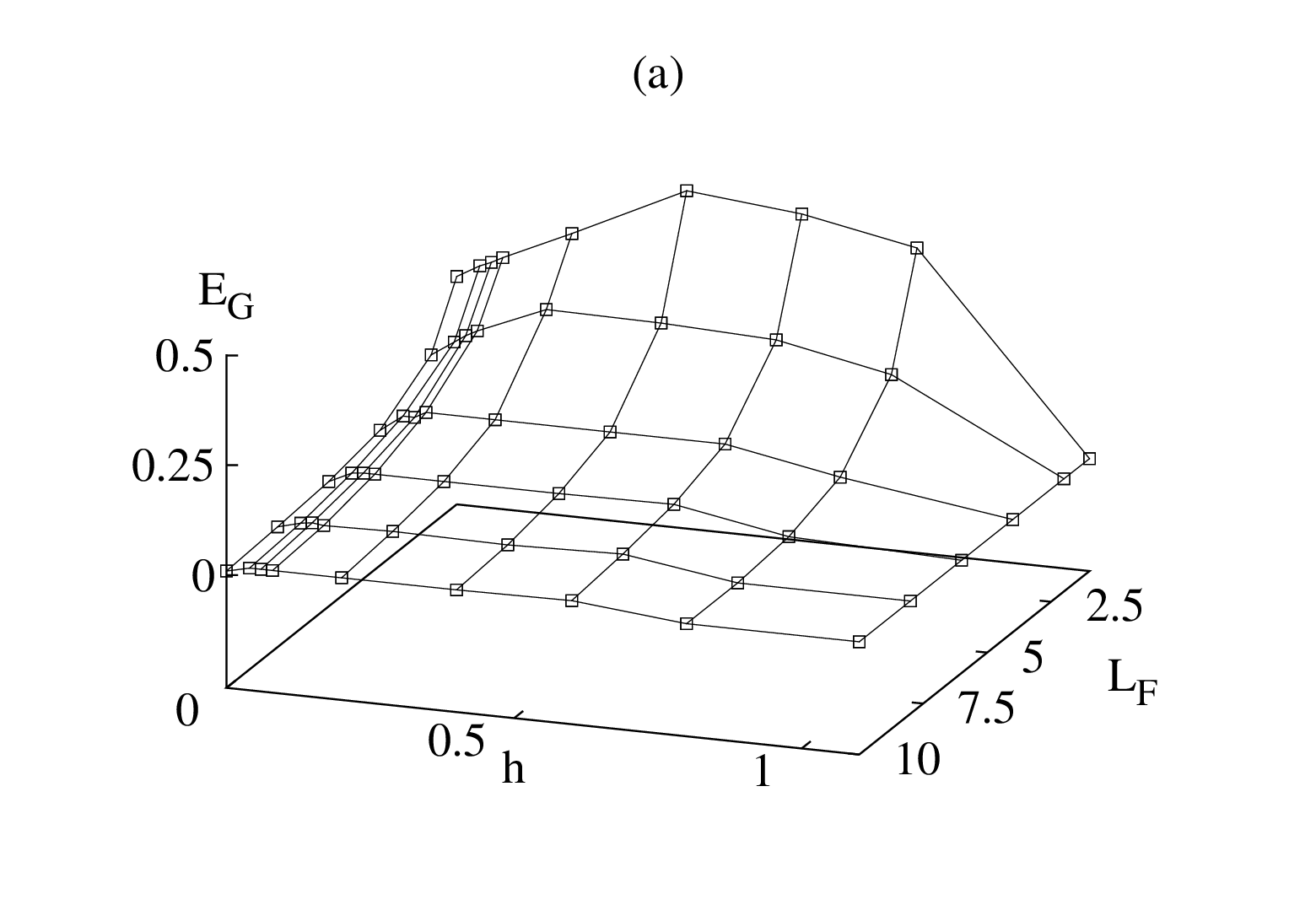}
\end{minipage}
\hspace{-1.6cm}
\begin{minipage}{4cm}
\epsfysize=6cm
\epsfxsize=8cm
\epsfbox{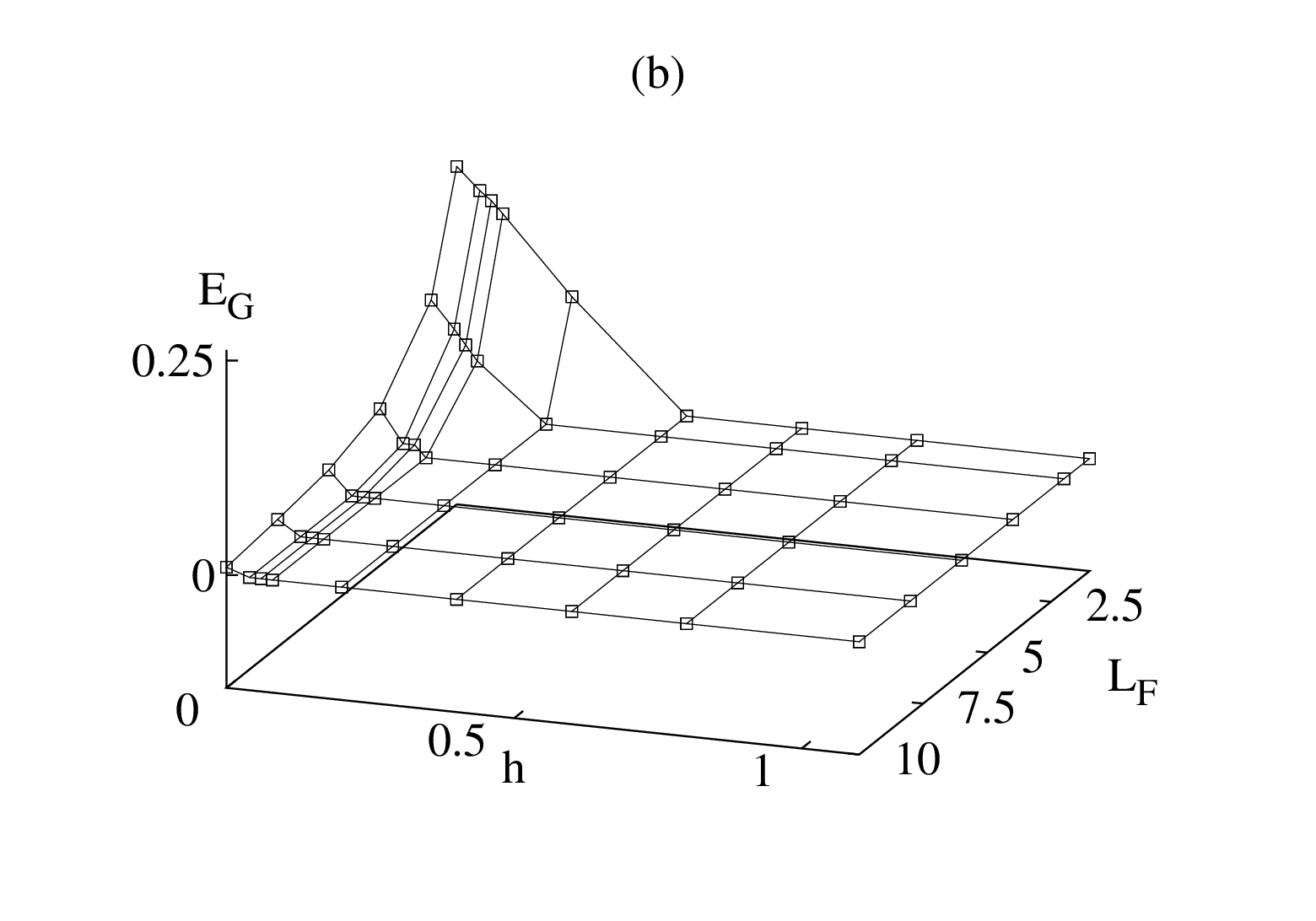}
\end{minipage}
}
\vspace*{-5.8cm} \hspace*{5cm}
\begin{large}
$
\uparrow
$
\hspace{6cm}
$
\downarrow
$
\end{large}

\vspace{5.8cm}
\caption{Behaviour of the gap as a function of length of the normal metal and the
        exchange field ($L_S =10$).
        ($\uparrow$) parallel spins, ($\downarrow$) antiparallel spins.}
\label{fig6}
\end{figure*}

The last issue we address is the behaviour of the DOS when the ferromagnet
has a finite length. Similarly to the SN system, a gap $E_{G}$ opens at
the Fermi energy. On the superconducting side, the gap shrinks
approaching the interface from the bulk. On the normal side,
the gap does not depend on the distance from the interface.
The gap location and width, however, do depend on  the spin direction and on 
the value of the exchange field. We only concentrate on the gap which opens at 
the Fermi energy.  For parallel spin $E_{G}$ is suppressed to zero when the 
(pair-breaking) exchange field is switched on (Fig.~\ref{fig6}a) whereas for 
antiparallel spin it has a  {\rm non-monotonous} behaviour (Fig.~\ref{fig6}b). 
The gap  first increases by increasing the exchange field up to $h \sim 0.5$, 
then, for higher values of the exchange field it is rapidly closed. 
For larger fields the position of the gap is roughly centered around 
$\mu_B H$ for spin down electrons. 
We postpone a more detailed description of the DOS of FS bilayers and  
multilayers (including the important role of spin-orbit scattering) 
to a forthcoming publication~\cite{future}  

In this Letter we studied the local DOS for a FS hybrid system. Most of the 
effects discussed here are pronounced if the  exchange field is of the order 
of the superconducting gap $\Delta_{BCS}$ like in YCo$_{2-x}$ Al$_{x}$ 
(which has a ferromagnetic transition around $10K$). A system closely 
related is a planar thin-film SN in a parallel magnetic field. Besides the 
fact that a Zeeman splitting is present in the superconductor, most of the 
results of this  remain valid and can be checked experimentally.

\acknowledgments
We thank F. Beltram, W. Belzig, C. Bruder, G. Falci, A. Filip, G. Giaquinta,
M. Giroud, F.W.J. Hekking, B. Pannetier and J. Siewert for  many useful 
discussions. 
The support of  the  PAIS-EISS of INFM is gratefully acknowledged.


\newpage

\begin{references}
\bibitem{deutscher69}
        G. Deutscher and P.G. de Gennes, in {\it Superconductivity} 
        R.D. Parks Ed. (M Dekker, NY, 1969).
\bibitem{reviews}C.W.J. Beenakker in {\em Mesoscopic
        Quantum Physics}, edited by E. Akkermans,  G. Montambaux,
        and J.-L Pichard (North Holland, Amsterdam) 1995;
        R. Raimondi and C.J. Lambert, J. Phys. C {\bf 10}, 901 (1998).
\bibitem{petrashov93}
        V.T. Petrashov, V.N. Antonov, P. Delsing, and, 
        T. Claeson, Phys. Rev. Lett. {\bf 70}, 347 (1993).
\bibitem{charlat96}
        P. Charlat, H. Courtois, Ph. Gandit, D. Mailly, and B. Pannetier, 
        Phys. Rev. Lett. {\bf 77}, 4950 (1996).
\bibitem{stoof96}
        T.H. Stoof and Yu.V. Nazarov,
        Phys. Rev. Lett. {\bf 76}, 2981 (1996).
\bibitem{gueron96}
        S. Gu\'eron, H. Pothier, N. O. Birge, D. Esteve,
        and M. Devoret, Phys. Rev. Lett. {\bf 77}, 3025 (1996).
\bibitem{golubov88}A. A. Golubov and M. Yu. Kupriyanov, J. Low Temp. 
        Phys. {\bf 70}, 83 (1988).
\bibitem{belzig96}
        W. Belzig, C. Bruder, and G. Sch\"on, Phys. Rev. B {\bf 54}, 
        9443 (1996).
\bibitem{frahm96} 
        K.M. Frahm, P.W. Brouwer, J.A. Melsen, and C.W.J. Beenakker, 
        Phys. Rev. Lett. {\bf 77}, 823 (1996).
\bibitem{altland98} A. Altland, B. D. Simons, D. Taras-Semchuk,
        JETP Lett. {\bf 67}, 1 (1998).
\bibitem{fazio97} 
        R. Fazio and G. Sch\"on in {\em Mesoscopic Electron Transport},
        NATO ASI series E, Vol, 345, p. 407, Kluver (1997).
\bibitem{bulaevskii85} 
        L.N. Bulaevskii, A.I. Buzdin, M.L. Kul\'ic, and S.V. Panyukov 
        Adv. Phys. {\bf 34}, 175 (1985).
\bibitem{radovic91}
        Z. Radovic, M. Ledvij, L. Dobrosavljevi\'c-Gruji\`c, A.I. Buzdin,
        and J.R. Clem, Phys. Rev. B {\bf 54}, 9443 (1996). 
\bibitem{demler97}
        E.A. Demler, G.B. Arnold, and M.R. Beasley,
        Phys. Rev. B {\bf 55}, 15174 (1997). 
\bibitem{dejong95} 
        M.D.M. de Jong and C.W.J. Beenakker, 
        Phys. Rev. Lett. {\bf 74}, 1657 (1995).
\bibitem{upadhyay98} S.K. Upadhyay, A. Palanisami, R.N. Louie, and R.A.
        Buhrman, Phys. Rev. Lett. {\bf 81}, 3247 (1998).
\bibitem{longrange}
        V.T. Petrashov, JETP Lett. {\bf 59}, 551 (1994);
        M.D. Lawrence and N. Giordano, J. Phys. C.{\bf 8}, 563 (1996).
\bibitem{giroud98}
        M. Giroud, H. Courtois, K. Hasselbach, D. Mailly and B. Pannetier, 
        Phys. Rev. B {\bf 58}, R11872 (1998).
\bibitem{leadbeater98}
        M. Leadbeater, C. J. Lambert, R. Raimondi, A. F. Volkov,
        cond-mat/9811117.
\bibitem{meservey94} R. Meservey and P.M. Tedrow, 
        Phys. Rep. {\bf 238}, 173 (1994). 
\bibitem{schmid81} 
        A. Schmid in {\em Nonequilibrium
        Superconductivity, Phonons, and Kapitza Boundaries}, 
        NATO ASI Series B 65,
        ed. K.E. Gray, (Plenum, New York 1981).
\bibitem{buzdin88} 
         A.I. Buzdin and L.N. Bulaevskii, 
        Sov. Phys. JETP {\bf 67}, 576 (1988).
\bibitem{tokuyasu88}
        T. Tokuyasu, J.A. Sauls, and D. Rainer,
        Phys. Rev. B {\bf 38}, 8823 (1988).
\bibitem{loff}
        A.I. Larkin and Yu. N. Ovchinnikov, 
        Sov. Phys. JETP {\bf 20}, 762 (1965);
        P. Fulde and R.A. Ferrell, 
        Phys. Rev. A {\bf 135}, 550 (1964).
\bibitem{future}        
        R. Fazio and C. Lucheroni, in preparation.
\end{references}
\end{document}